 \documentclass[final,5p,times,twocolumn]{elsarticle}

\usepackage{amssymb}
\usepackage{color}
\usepackage{amsmath}
\usepackage{url}
\usepackage{graphics}

\biboptions{sort&compress}

\journal{Nuclear Instruments and Methods in Physics Research A}

\begin{document}

\begin{frontmatter}

%% Title, authors and addresses

%% use the tnoteref command within \title for footnotes;
%% use the tnotetext command for the associated footnote;
%% use the fnref command within \author or \address for footnotes;
%% use the fntext command for the associated footnote;
%% use the corref command within \author for corresponding author footnotes;
%% use the cortext command for the associated footnote;
%% use the ead command for the email address,
%% and the form \ead[url] for the home page:
%%
%% \title{Title\tnoteref{label1}}
%% \tnotetext[label1]{}
%% \author{Name\corref{cor1}\fnref{label2}}
%% \ead{email address}
%% \ead[url]{home page}
%% \fntext[label2]{}
%% \cortext[cor1]{}
%% \address{Address\fnref{label3}}
%% \fntext[label3]{}

\title{Load management strategy for Particle-In-Cell simulations in high energy particle acceleration}
%% use optional labels to link authors explicitly to addresses:
%% \author[label1,label2]{<author name>}
%% \address[label1]{<address>}
%% \address[label2]{<address>}

\author[llr]{A. Beck\corref{cor1}}\ead{beck@llr.in2p3.fr}
\author[nbi]{J. T. Frederiksen}
\author[cea]{J. D\'{e}rouillat}

\cortext[cor1]{Corresponding author.}
\address[llr]{Laboratoire Leprince-Ringuet, \'{E}cole Polytechnique, CNRS-IN2P3, Palaiseau 91128, France.}
\address[nbi]{Niels Bohr Institute, University of Copenhagen, Blegdamsvej 17, 2100 K{\o}benhavn {\O}, Denmark}
\address[cea]{CEA, Maison de La Simulation 91400 Saclay, France.}

\begin{abstract}
In the wake of the intense effort made for the experimental CILEX project, numerical simulation campaigns have been carried out in order to finalize the design
of the facility and to identify optimal laser and plasma parameters. These simulations bring, of course, important insight into the fundamental physics at play.
As a by-product, they also characterize the quality of our theoretical and numerical models. 
In this paper, we compare the results given by different codes and
point out algorithmic limitations both in terms of physical accuracy and computational performances.
These limitations are illustrated in the context of electron \textit{l}aser \textit{w}ake\textit{f}ield \textit{a}cceleration (LWFA).
The main limitation we identify in state-of-the-art \textit{P}article-\textit{I}n-\textit{C}ell (PIC) codes is computational load imbalance. We propose an innovative
algorithm to deal with this specific issue as well as milestones towards a modern, accurate high-performance PIC code for high energy particle acceleration.
\end{abstract}

\begin{keyword}
%% keywords here, in the form: keyword \sep keyword
Laser wakefield acceleration \sep petawatt laser \sep bubble regime \sep electron self-injection \sep relativistic self-focusing \sep  particle-in-cell simulations
\sep Photon-Plasma \sep SMILEI \sep dynamic load balancing
%% MSC codes here, in the form: \MSC code \sep code
%% or \MSC[2008] code \sep code (2000 is the default)
%\PACS 41.75.Jv \sep 52.38.-r \sep 52.35.Mw \sep 52.27.Ny \sep 52.65.Rr \sep  52.65.Ww
\end{keyword}

\end{frontmatter}

%%
%% Start line numbering here if you want
%%
%\linenumbers

%% main text
\section{Introduction}
\label{sec-intro}

%text from Serguei
% Acceleration of electrons in laser-driven plasma wakes has been the focus of keen interest over the last three decades. The introduction of laser facilities delivering
% 100-terawatt (TW)-class mid-IR pulses at high repetition rates greatly advanced this area. With such lasers, it became possible to achieve a complete ponderomotive
% blowout of the electron fluid in low-density plasmas (keeping the much heavier ions at rest), maintaining the cavity (``bubble'') of electron density over many
% Rayleigh lengths. The resulting extension of the acceleration length to the centimeter range, combined with a GeV/cm-scale gradient supported by the bubble, culminated
% in production of low phase-space-volume electron bunches with energies close to 1 GeV \cite{Nakamura_GeV_2007,Banerjee_GeV_2012,JLee_GeV_2013}. A sea change in the 
% field of laser plasma acceleration (LPA) should come with the further progress in laser technology. The state-of-art systems are presently delivering short (15--30 fs
% petawatt (PW) laser pulses with 0.1--10 Hz repetition rate \cite{JLee_GeV_2013,Aoyama_PW_2003,JLee_PW_2010,Umstadter_PW_2013}. Per standard LPA scalings
% \cite{Lu_2007PhRvS}, experiments at these facilities promise boosting electron energy to a few GeV, producing beams with parameters competitive with the standard
% linacs, but with the flexibility in parameters not easily afforded by the standard accelerator technology.

The future CILEX (Centre Interdisciplinaire de la Lumiere EXtr\^{e}me/Interdisciplinary Center for the Extreme Light) is a state-of-the-art laser facility that targets
diverse applications using plasmas produced by short (15--30 fs) multi-petawatt (1-10 PW) laser pulses with 0.1-10 Hz repetition rate. It will host the Apollon-10P laser, 
which will deliver pulses with an instantaneous power up to 10 PW, and the associated infrastructures and experimental setups.
It will thus offer an opportunity for scientific breakthrough in various domains \cite{Cros_2013NIMA}.
The facility is located in France (Paris area) and is expected to open to the international user community in 2016.

The experimental setups and the required instruments for relevant diagnostics remain to be designed.
New regions of the parameter space are now accessible with Apollon-10P.
They have never been reached before and large numerical simulation campaigns are necessary to explore them.
 These campaigns have three objectives:
 (1) 
 %To finalize the design of the CILEX facility by 
 to make sure that the most promising laser and plasma parameters are accessible in CILEX,
 (2) 
 %to evaluate the outcome of the experiments 
 to improve the preparation in terms of instrumentation and radioprotection,
 and (3) to help understanding the physics at work.
 %, in these new regimes.

Computational support is therefore mandatory for such large experimental facilities, and PIC simulations are the best candidates for this task.
A large variety of PIC codes exists.
They range from simple serial codes making severe physical assumptions to speed up computations, to elaborate and computationally costly massively parallel PIC codes, supposedly more accurate. 
%To put it simply, there is a trade-off between accuracy and cost of the simulation. 

In Section \ref{sec:Sec2}, we present a first numerical study of a typical CILEX case of LWFA using three standard but very different codes, to
understand their strengths and limitations. The LWFA is at the core of the CILEX scientific program, with the principle task being to explore
the experimental prospects of using ultra-short PW pulses for multi-GeV acceleration;
this will occur in the bubble regime \cite{Pukhov_2002}, taking advantage of electron self-injection and laser pulse self-guiding in a low-pressure gas/plasma cell.
Section \ref{sec:Sec3} discusses computational load imbalance, how it heavily slows down LWFA simulations but
also how this barrier can be lifted with an appropriate dynamic load balancing algorithm.

\section{Self injection in the bubble-out regime at CILEX}
\label{sec:Sec2}
\subsection{The codes}
This section details the results given by the PIC codes Wake, Calder-Circ (CC) and Photon-Plasma (PP) for a standard CILEX LWFA experiment.

%WAKE and previous application
 Wake is a quasi-static, axi-symmetric code \cite{Mora_1997PhPl} with
test particle tracking \cite{2010NJPhKalmykov,2011PPCFKalmykov,Kalmykov_PoP_2011,kalmykov_NJP_2012}.
In spite of its restrictive context, Wake can be used to identify the nonlinear optical 
processes at work in the laser-plasma interaction and have been used, for instance, to explain the details of pulse evolution for acceleration beyond the theoretical
pulse depletion limit\cite{Beck_2013NIMA}.
Wake is a sequential code and a typical LWFA simulation only takes half a day of computation on a single core.

%Calder-Circ and previous application 
CC is an explicit quasi-3D PIC code \cite{Lifschitz_2009JcP}. It uses poloidal mode decomposition for the electromagnetic fields and currents, 
while computing super-particles (SP) trajectories in the 3D Cartesian space. Using the two lowest-order modes is sufficient to accurately capture the most important aspects of the
interaction \cite{Lifschitz_2009JcP}. CC has been used in many laser-plasma interaction studies where long laser propagation (a few cms) were needed\cite{Beck_2013NIMA,2010NJPhKalmykov,Cowan_JPP_2012}.
Indeed, the mode decomposition allows to get 3D results for the cost of a couple of 2D standard simulations and is well adapted to long propagations. A typical LWFA simulation
with CC takes several days of computation on 500 cores.   

%Photon-plasma and novelties
PP is an explicit fully 3D PIC code. It benefits from high-order schemes and has demonstrated excellent scalability up to 250,000+ cores using a hybrid MPI+OpenMP approach\cite{Haugboelle_2013}.
It has been used mostly in astrophysical cases\cite{Frederiksen_2008} and was recently upgraded to support long laser propagations.
For this application, its high-order scheme is especially valuable.
With a 6th order scheme, one can conserve stability and suppress numerical dispersion (for a given wavelength) by simply setting the appropriate resolution in time and space.
In contrast, numerical dispersion in a traditional 2nd order scheme cannot be reduced without improving temporal and spatial resolution, which is orders of magnitude more
costly.
Another benefit of the sixth order scheme is that the numerical dispersion, when not suppressed,
tends to overestimate the group velocity of the laser and therefore limits the numerical Cherenkov effect with respect to lower order schemes who tends to underestimate the
laser group velocity.

\subsection{Simulation parameters}
The simulation presented here uses the expected laser parameters of the first CILEX shots with the so called ``F2'' beam. The plasma parameters have been optimized
as shown in \cite{Beck_2013NIMA}
and with a complementary parametric study with Wake. They provide a good trade off between electron energy, energy spread and total charge of the electron bunch.
The carrier wavelength of the  Apollon-10P laser is $\lambda_0 \! = \! 0.8$ $\mu$m. The laser pulse is Gaussian in all directions has
a duration of $\tau_L\! = \! 25$ fs (FWHM in intensity),
the laser strength parameter is $a_0 \! = \! 5.55$ and its transverse spot size is $w_0\! = \! 16.4\: \mu m$.
The pulse energy on target is 7.41 J for a peak power of 0.28 PW. 
And finally, the plasma density is set to $n_e\!=\!1.4\times 10^{18}$ cm$^{-3}$ starting by a 0.3 mm long ramp and the laser is focused at the foot of the ramp.
%Using these parameters, we give a comprehensive analysis of the laser plasma interaction, including the associated relativistic optical phenomena.
%We focus on acceleration beyond 1 GeV, exploring the regimes that maximize the electron energy without beam quality degradation.
%A very basic configuration will be implemented in the first round of experiments.

Similar numerical parameters were chosen for PP and CC for benchmarking purposes.
The longitudinal and transverse resolutions in both codes are respectively set to $\Delta z\! =\! 0.125\, [c/\omega_0]$ and $\Delta x,y\! =\! 1.5\, [c/\omega_0]$ ,
where $\omega_0$ is the laser frequency. 
The time steps $\Delta t$ differ for stability reasons and are respectively 0.0625 and 0.124 $[1/\omega_0]$ for PP and CC.
Wake uses the coarser resolution $\Delta z\! =\! 0.47\, [c/\omega_0]$, $\Delta x\! =\! 5.55\, [c/\omega_0]$ and $\Delta t\!= 15.42\, [1/\omega_0]$.

%The results of three-dimensional PIC simulations reported here show that few-GeV electron bunches can be produced with a low phase space volume
%and high enough charge to be used for either radiation physics applications or staged acceleration; both applications are the top priorities on the CILEX
%  agenda \cite{Cros_2013NIMA}.

\subsection{Results}

\begin{figure}
\centering
\includegraphics{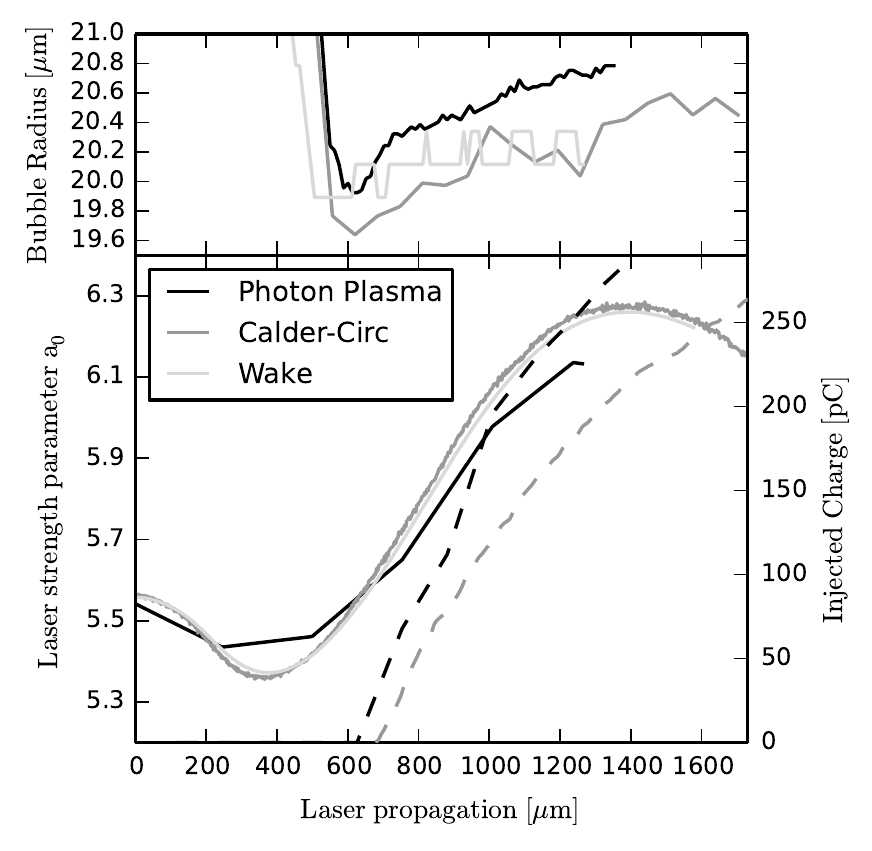}
\caption{Evolution of the amplitude of the driver, injected charge and bubble radius given by the three different codes.
On the bottom panel, plain lines show the evolution of $a_0$ and dashed lines the injected charge.
Wake does not give any information about the injected charge because it assumes a quasi-static evolution.}
\label{fig:fig1}
\end{figure}

The simulations were run long enough to give a quantitative evaluation of the entire injection process at play.
The most important results are summarized in Figure \ref{fig:fig1}.
% Two specific setups have been investigated. F2, the secondary, less intense laser beam of 250 TW, and F1, the primary beam with an initial expected peak power of 1.6
% PW. These are the expected characteristics of the two beams at the launch of the Cilex facility next year.
After the entry of the laser pulse in the plasma density ramp, the laser self-focuses and the so-called
bubble structure forms in its wake.  The bubble quickly shrinks to its minimum size before starting to grow again very slowly as injection begins.
% In the F2 case for instance, self focusing starts after $\rm 400\ \mu m$ of propagation and stops around $\rm 1250 \mu m$. During this period of time, the bubble
% radius decreases quickly from $\rm 25\ \mu m$ to below $\rm 20\ \mu m$ at $600\ \mu m$ of propagation and then slowly grows again to reach $\rm 20.6\ \mu m$ after
% $\rm 1400\ \mu m$ of propagation. Electron injection occurs during the ``growing phase''. 275 pC of electrons are injected after $\rm 1400 \mu m$ of propagation.
% These clean results will impact decisions in the final design of the Cilex facility. It is a good demonstration that we need to work with lower density plasmas, make
% sure that the laser is capable of producing sufficiently wide pulses and that peak intensity is not necessarily the priority if one wants to produce clean,
% monoenergetic electron beams.
The three codes show a similar scenario, with, in particular, a similar relativistic self-focusing of the pulse. However, even though the driver is well described
in all codes, the bubble dynamics and the total charge of injected electrons differ. After 1400 $\rm \mu m$ of propagation,
275 pC are injected in PP whereas only 200 pC are injected in CC. 
This 30\% difference is pretty small with respect to the variations observed experimentally from shot to shot.
%Being able to evaluate the injected charge accurately is extremely important and is one of the current challenges faced by the laser-plasma electron acceleration community.
It could be explained by the fact that
%points toward possible improvements to the CC code. In PP the simulation is fully 3D and the injection process, occurring
%on axis, is well described. 
CC
%, on the other hand,
is based on a cylindrical geometry and the longitudinal axis is subject to user-defined boundary conditions
which might not always be accurate and could artificially influence the injection. It is also known that the numerical Cherenkov effect introduces a significant amount of noise
on axis in CC\cite{Lehe_2013}.
% Simulations of the same physical cases were performed on Curie Thin Nodes using the Calder-Circ code. This code assumes a cylindrical symmetry  and therefore, has a
% computational load equivalent to a two dimensional simulation but is supposed to give a good description of the 3D process. The sequential code Wake was also used to 
% simulate the same cases. Wake makes additional physical assumptions (quasi-static) to simplify the set of equations solved but is unable to describe the behavior of
% the injected electron beam. Nevertheless, it gives an accurate description of the laser pulse propagation. The comparison between the three simulations of the same
% physical cases gives very important insights to the numerical scientists. It illustrates exactly which part of the physical process is missing after each additional
% simplification. It is also an important benchmark, validating, or not, the latest developments.
% For both cases F1 and F2, Calder-Circ and Photon-Plasma  give very similar dynamics. In particular, similar self-focusing and plasma responses.
% The laser initialization and the innovative 6th order scheme in Photon-Plasma is therefore fully validated. 
Nevertheless, even though here CC gives results very similar to PP, full 3D simulations would still be mandatory in cases where the cylindrical symmetry is broken such as
laser pulses with experimental aberrations, the presence of an external magnetic field or a transverse density gradient.

%\section{The importance of being load-balanced}
\section{The importance of computational load-management}
\label{sec:Sec3}

\begin{figure}
\centering
\includegraphics[height=0.35\textheight]{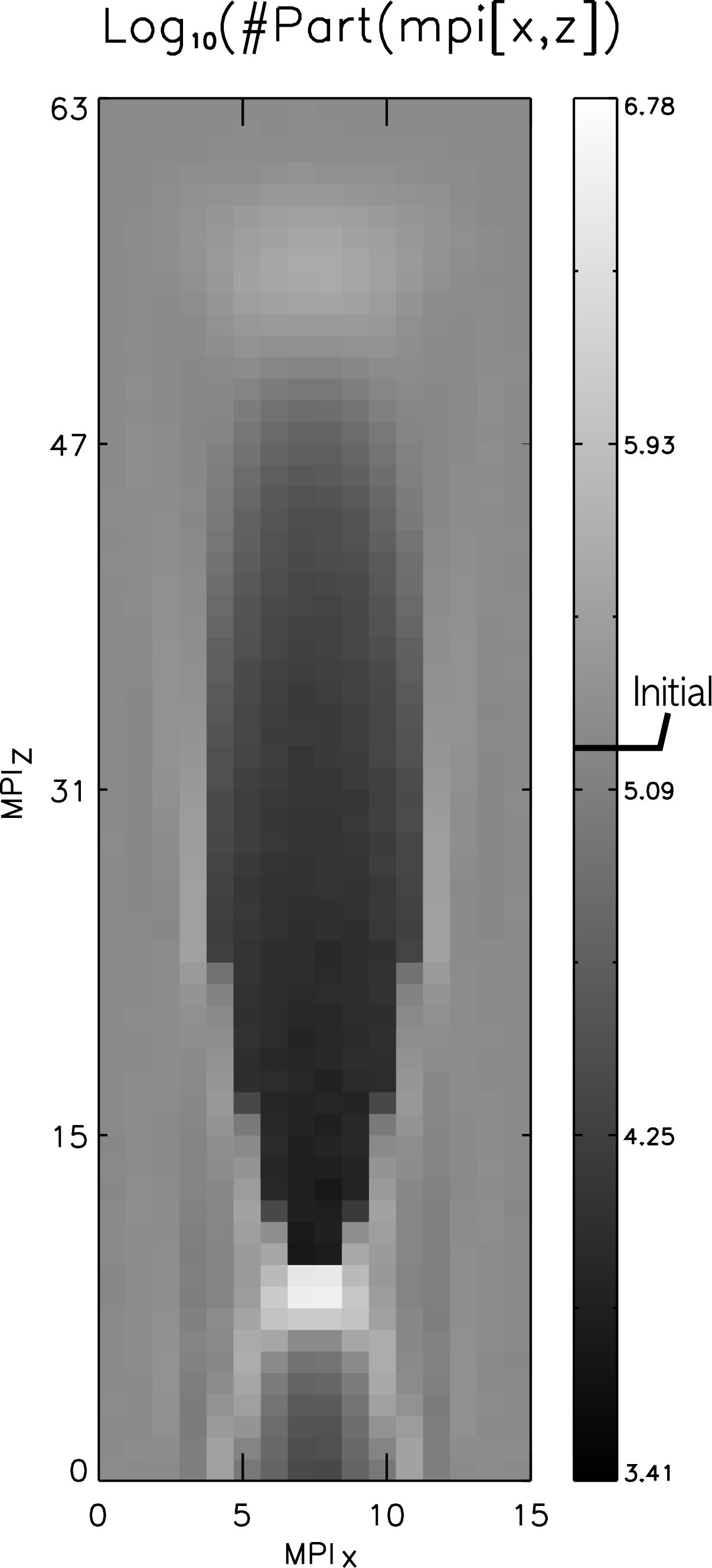}
\caption{ Total number of super-particles owned by each MPI process in a full 3D Photon-Plasma simulation of LWFA. It ran on 2048 nodes of the Fermi system at Cineca. Note the base-10 logarithmic scaling. 
The initial value is Log$_{10}$(N$_{init,MPI}$) = 5.19 --- or, equivalently, N$_{init,MPI}$ = 155 520. The minimum and maximum are N$_{min,MPI}$ = 2570,
and N$_{max,MPI}$ = 6 070 158, respectively.}
\label{fig:fig7}
\end{figure}

Fully 3D PIC simulations are rarely used and only by groups with massive access to top-tier supercomputers.
Their high cost is often prohibitive.
In some cases, the raw computational cost can be significantly reduced by the use of a Lorentz boosted frame\cite{Vay_2007}. But even in these favorable cases, 
they still require a very large number of cells and SP and fit only on massively parallel systems.
%This is actualy not very selective since nowadays, many groups
%have access to systems capable of running 3D simulations. But things become complicated when load imbalance sets in.
It is almost inevitable that, at some point of the run, a small part of the system becomes overloaded with too many SP, slowing down the entire simulation.
At the same time, a large part of the system is underloaded and remains idle, wasting precious computational resources.
This phenomenon, know as load imbalance, is more likely to happen as the size of the system and the number of spatial dimensions increases. 
Massively parallel 3D cases being most problematic as seen in Figure \ref{fig:fig7}. 
This very large simulation is a remarkable illustration of the limitation of actual 3D LWFA simulations.
Some areas of the simulation are almost completely empty of SP whereas some are very densely populated and handle 40 times as many SP as they did in the initially uniform plasma.
A slow-down by a similar factor is therefore expected, and may lead to unacceptable under-utilization of the computational resources.

The use of OpenMP is often believed to be a solution to load imbalance in PIC codes.
More OpenMP threads implies fewer and larger MPI domains and consequently smoother SP distribution between them.
Since the load is directly related to the number of SP, it is also shared more homogeneously. 
%OpenMP threads, effectively decreasing load imbalance between MPI domains. 
Nevertheless, in the case of Figure \ref{fig:fig7}, 16 OpenMP threads per MPI process were used and yet a very strong imbalance remains.
The openMP effectiveness is limited on large systems because as the number of cores increases, the number of OpenMP threads remains the same, bounded by the number of hardware threads per CPU.
As a consequence, the number of MPI processes must increase and the relative sizes of the MPI domains shrink, eventually becoming smaller than the load variation scale.
At this point, load imbalance hits and  performances are lost. OpenMP can delay this situation, but not prevent it.
Consequently, implementation of some form of dynamic load balancing to achieve scalability, is paramount.
Below, two innovative approaches are considered: dynamic load balancing via ``patching'' and dynamic load limiting via k-means clustering.
These methods are not mutually exclusive, rather, they are complementary.
%-- nor unilaterally -- 
%exclusive.
 
 \subsection{Dynamic load balancing}
The ``patching'' method was first described in \cite{Germaschewski_2014} and has been implemented in SMILEI \cite{Smilei}, a PIC code developed to support the CILEX community.
It consists in breaking the large structures encapsulating fields and SP data of each MPI process into a multitude of smaller independent structures called ``patches''.
%Being independent, these patches require an additional synchronization step but with a very limited cost
%because most of the synchronizations occur between patches handled by the same node and do not need communications.
The benefit is that the code now has a data structure particularly well adapted to thread parallelization and, in particular,
to OpenMP.

\begin{figure}
\centering
\includegraphics[width=0.8\columnwidth]{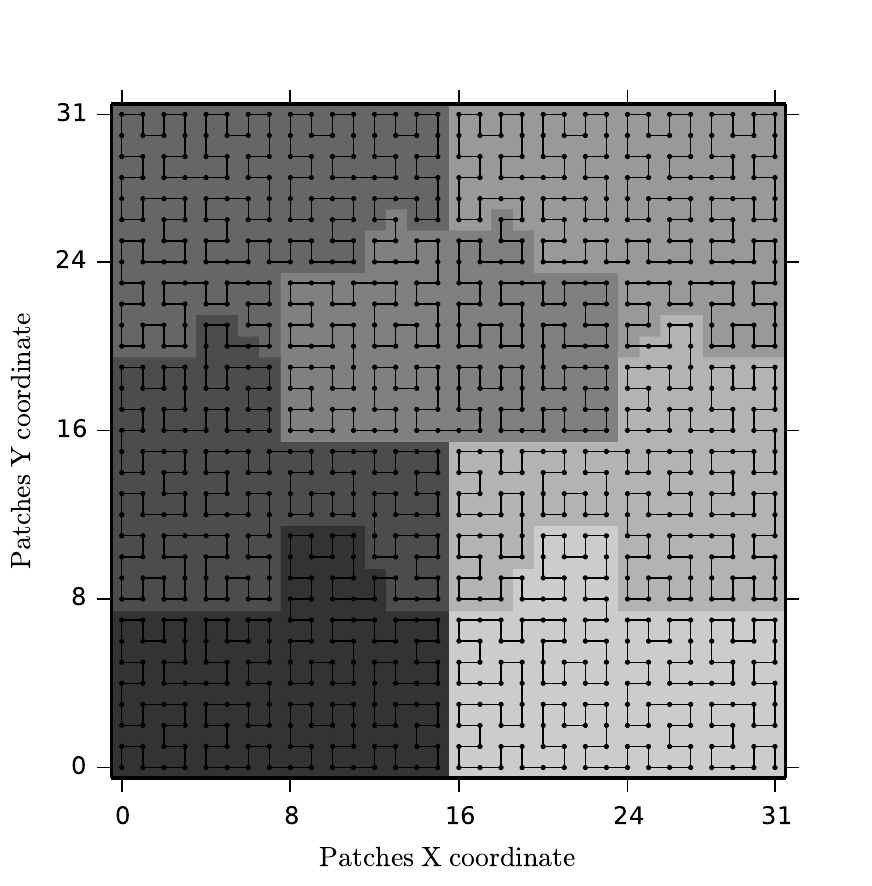}
\caption{Example of a $32\times32$ patches decomposition between 7 MPI processes. MPI domains are delimited by different uniform colors. The line shows the Hilbert
curve and the dots the center of the patches. The curve goes through all patches. It starts from the patch with coordinates $(0,0)$ and end at patch with coordinates
$(31,0)$.}
\label{fig:fig2}
\end{figure}

Patches are coherently organized along a 1D Hilbert space filling curve\cite{Hilbert_1891}. Each MPI process owns a segment of this curve. 
Its properties of continuity and locality ensure that the obtained domains are compact and minimize synchronization communications.
This decomposition has a very flexible organization and can be very different from a Cartesian decomposition as shown in Figure \ref{fig:fig2}.
Taking advantage of this flexibility, MPI processes are able to dynamically exchange patches  in order to even out their respective loads.

\begin{figure}
\centering
\includegraphics[width=0.8\columnwidth]{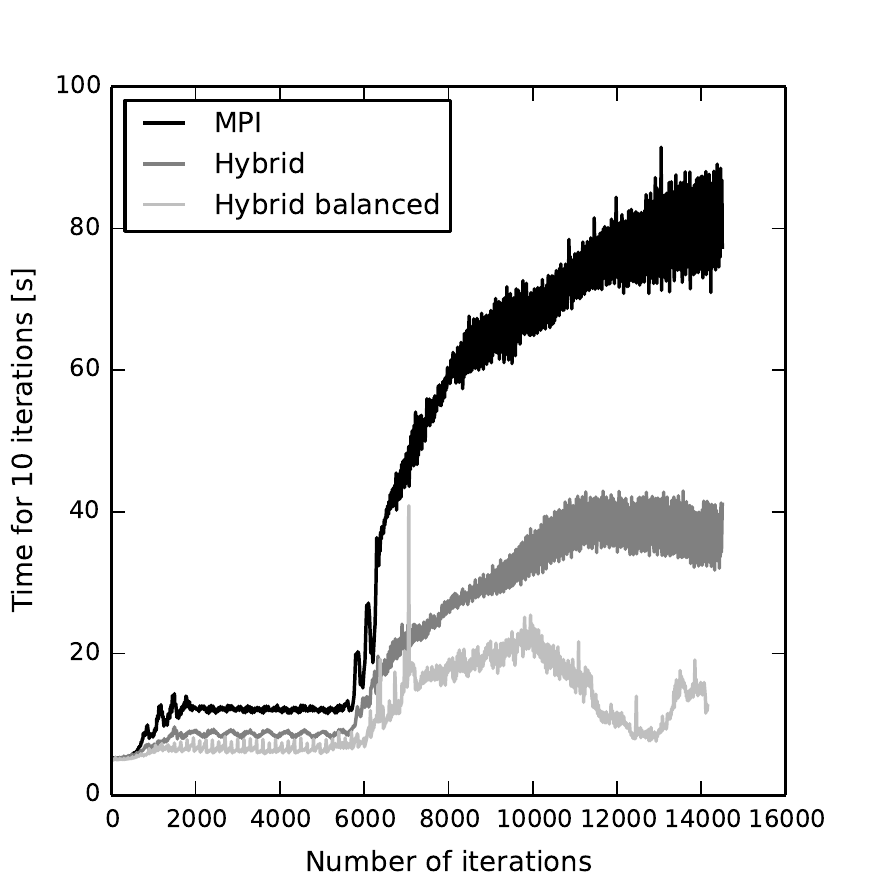}
\caption{Evolution of the time spent for 10 SMILEI cycles along a LWFA run on 24 OCCIGEN nodes. ``MPI'' is a pure MPI run, 576 MPI processes.
``Hybrid'' is a hybrid MPI+OpenMP run, 48 MPI processes + 12 OpenMP threads per process. ``Hybrid balanced'' is the same hybrid run
with the dynamic load balancing active.}
\label{fig:fig4}
\end{figure}

Results for a simple 2D LFWA simulation with SMILEI running on 24 nodes of the OCCIGEN system are compared in Figure \ref{fig:fig4}.
As shown in \cite{Germaschewski_2014}, load imbalance
occurs early in the simulation. In the LWFA case, this sets in already at bubble formation around iteration 6000. 
The first few hundreds iterations take around 0.5 s each, independent of  parallelization strategy.

In a run with a pure MPI parallelization, the time needed for 10 iterations increases up to 80 s which gives an imbalance ratio of approximately 14.
In the hybrid run with 12 OpenMP threads, this ratio is approximately halved and is only around 2 when the dynamic load balancing is activated.

 \subsection{Dynamic load limiting}

The other approach is based on computational SP splitting/merging in low/high SP number density cells, respectively.
It has been implemented in PP. 
The merging/splitting must carefully take into consideration conservation laws of physics; the scheme is costly, but precise.
It is based on an accelerated k-means clustering method.
Besides being of importance in load imbalanced situations, merging and splitting of SP is of profound significance when carrying out simulations in which detailed Monte Carlo
particle-particle collisions, ionization or quantum electrodynamics (QED) are implemented. 
Otherwise, they can lead to an excessive number of SP and to memory bound situations because of the tremendous amount of new SP created; 
smaller SP of the same species from collisions, electrons in the case of ionization and electron-positron pairs or even photons from QED.
 In this regard, this approach can be considered more as a dynamic load limiting algorithm rather than balancing. 
It is absolutely mandatory for modern simulations since these new kind of developments are emerging in both laboratory and astrophysical plasmas communities.
Details of this approach can be found in \cite{Frederiksen_2015}. 

\section{Summary and outlook}

The importance of 3D simulations has been established. Their cost is still prohibitive for most applications since large scale simulations
are so massively parallel that load imbalance becomes a barrier.
It cannot be alleviated by brute force parallelism. That is merely a waste of computational resources.
Germaschewski et al. suggested a new data structure for PIC codes, adapted to both modern massively parallel hardware and to the associated algorithmic challenges \cite{Germaschewski_2014}.
This structure has been adapted with success in SMILEI \cite{Smilei} and shows tremendous improvement in the case of LWFA, even in 2D.

Recent advances in the physics included in PIC codes in both astrophysical plasmas and laser-plasma communities often leads to prohibitive numbers of SP.
 Now, the problem is not just only compute bound, but also memory bound.
 This issue may be resolved only through a load-limiting algorithm such as particle merging through k-means clustering \cite{Frederiksen_2015}.

The two methods above could be combined into a coherent dynamic load management strategy without interfering with any of the physics included in the code. It is even compatible
with different geometry models like CC itslef which is also subject to load imbalance.

\section*{Acknowledgments}

PP and CC simulations 
were performed using the PRACE Research Infrastructures
resources FERMI based in Italiy at CINECA and Curie based in France at TGCC under the grant 2013091859. SMILEI simulations were performed using DARI allocation c2015067484 on the
OCCIGEN system, based in France at CINES. This work was partially supported with a bi-lateral STSM visit grant under the EU COST Action MP 1208 ``Developing the physics
and the scientific community for inertial fusion'' and by the LabEx PALM (ANR-10-LABX-0039-PALM) under project SimPLE. 
%
%% References with bibTeX database:

%\bibliographystyle{model1-num-names}
\bibliographystyle{elsarticle-num}

%\bibliography{EAAC_2013_proceedings.bib}

%% Authors are advised to submit their bibtex database files. They are
%% requested to list a bibtex style file in the manuscript if they do
%% not want to use model1-num-names.bst.

%% References without bibTeX database:

% \begin{thebibliography}{00}

%% \bibitem must have the following form:
%%   \bibitem{key}...
%%

% \bibitem{}

% \end{thebibliography}

\end{document}